# An automated system for lung nodule detection in low-dose computed tomography

I. Gori*[a,b], M. E. Fantacci[b,c], A. Preite Martinez[d], A. Retico[b]
[a]Bracco Imaging S.p.A., Milano, Italy
[b]Istituto Nazionale di Fisica Nucleare, Sezione di Pisa, Italy
[c]Dipartimento di Fisica, Università di Pisa, Italy
[d]Centro Studi e Ricerche Enrico Fermi, Roma, Italy

## ABSTRACT

A computer-aided detection (CAD) system for the identification of pulmonary nodules in low-dose multi-detector helical Computed Tomography (CT) images was developed in the framework of the MAGIC-5 Italian project. One of the main goals of this project is to build a distributed database of lung CT scans in order to enable automated image analysis through a data and cpu GRID infrastructure.

The basic modules of our lung-CAD system, a dot-enhancement filter for nodule candidate selection and a neural classifier for false-positive finding reduction, are described. The system was designed and tested for both internal and sub-pleural nodules. The results obtained on the collected database of low-dose thin-slice CT scans are shown in terms of free response receiver operating characteristic (FROC) curves and discussed.

**Keywords:** Methods: pre-processing, feature extraction, classification, evaluation and validation; Modalities: X-ray CT; Diagnostic task: detection

## 1. INTRODUCTION

Lung cancer is one of the most relevant public health issues. Despite significant research efforts and advances in the understanding of tumour biology, there was no reduction of the mortality over the last decades.

Lung cancer most commonly manifests itself with the formation of non-calcified pulmonary nodules. Computed Tomography (CT) is the best imaging modality for the detection of small pulmonary nodules, particularly since the introduction of the helical technology[1]. However, the amount of data that need to be interpreted in CT examinations can be very large, especially when multi-detector helical CT and thin collimation are used, thus generating up to about 300 two-dimensional images per scan, corresponding to about 150 MB. In order to support radiologists in the identification of early-stage pathological objects, researchers have recently begun to explore computer-aided detection (CAD) methods in this area.

Among the approaches that are being tried to reduce the mortality of lung cancer is the implementation of screening programs for the subsample of the population with higher risk of developing the disease. The First Italian Randomized Controlled Trial (ITALUNG-CT) that aims to study the potential impact of screening on a high-risk population using low-dose helical CT was recently started[2].

A CAD system for small pulmonary nodule identification, based on the analysis of images acquired from the Pisa centre of the ITALUNG-CT trial, was developed in the framework of the MAGIC-5 collaboration funded by Istituto Nazionale di Fisica Nucleare (INFN) and Ministero dell'Università e della Ricerca (MIUR). The system is based on a dot-enhancement filter for the identification of nodule candidates and a neural network based classification module for the reduction of the number of false-positive (FP) findings per scan.

*ilariagori@gmail.com; phone +39 340 6220624

## 2. MATERIAL AND METHODS

### 2.1 The lung CT database

A low-dose lung CT dataset was acquired from the Pisa centre of the ITALUNG-CT trial, the First Italian Randomized Controlled Trial for the screening of lung cancer[2]. The CT scans are acquired with a 4 slices spiral CT scanner according to a low-dose protocol (screening setting: 140 kV, 20 mA), with a 1.25 mm slice collimation.

The dataset used for this study consists of 39 low-dose CT scans. Each scan is a sequence of slices stored in DICOM (Digital Imaging and COmmunications in Medicine) format. The average number of slices per scan is about 300 with 512×512 pixel matrix, a pixel size ranging from 0.53 to 0.74 mm and 12 bit grey levels. The scans were annotated by experienced radiologists, by using a visualization and annotation tool we developed, described in the following sub-section.

Non-calcified solid nodules only, with a diameter greater than 5 mm, were considered in this study, whereas ground-glass opacities were excluded. The dataset consists of 102 annotated nodules, 75 internal nodules belonging to 34 out of the 39 available scans and 27 sub-pleural nodules belonging to 20 out of the 39 available scans.

Examples of internal and sub-pleural nodules extracted from our screening database are shown in Fig. 1.

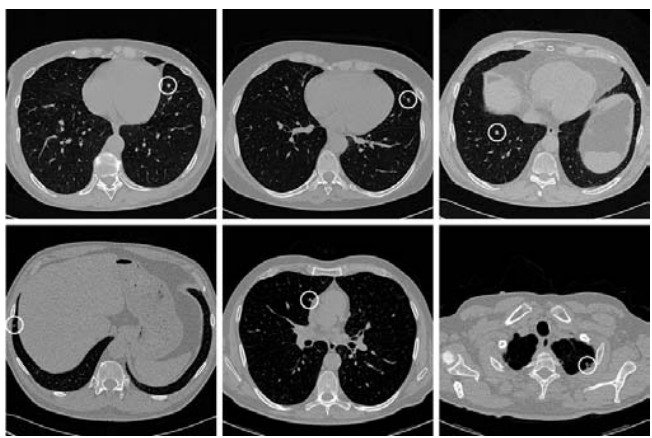

Fig. 1. Examples of internal (up) and sub-pleural (down) small pulmonary nodules.

### 2.2 The annotation tool

In order to be able to insert into our database the annotations required to precisely identify the position of every diagnosed nodule, we had to develop an *annotation tool*, i.e. a software application with a graphical user interface (GUI) where the radiologist can explore the CT data, identify eventual nodules and annotate them using a standard classification scheme. This tool is built upon the MG framework, and thus strongly integrated with the ROOT[3] platform, allowing access to GRID-enabled distributed resources. The tool could possibly evolve into the main interface for the complete lung CAD system.

The annotation tool can read three-dimensional CT data scans stored in DICOM format, and will read and write annotation lists in a simple text-based format. At the moment, since this tool was developed in order to validate the results of the lung CAD algorithms, an annotation is simply associated to a spherical region of the dataset, identified by the position of the center and its radius. This is all the information currently required by the CAD validation tools. If different requirements shall arise, the modular organization of the software will allow other selection methods to be implemented (manual or computer-assisted boundary identification, for example).

The graphical interface allows the radiologist to visualize axial slices of the CT data (the standard visualization mode that radiologists are used to) (see Fig. 2). Standard imaging and navigational controls are provided, such as zoom/pan functionality and dynamic range control (the so-called *window* selection).

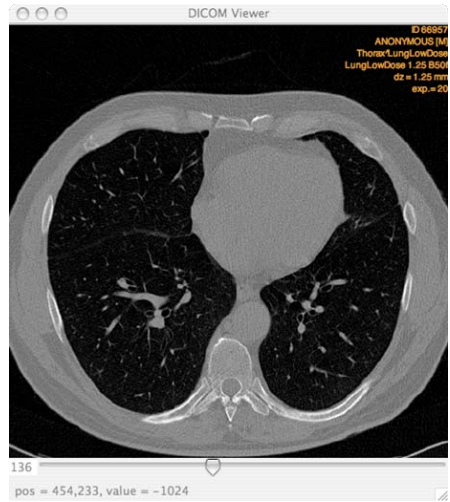

Fig. 2. Standard visualization mode of a single CT slice.

Once the radiologist identifies a suspect nodule, they can select that region using a click-and-drag interface: a pop-up window allows classification of the nodule according to a standard set of parameters (morphological, and eventually clinical). The selected region may be further examined with the aid of orthogonal cross-sections (in standard and *Maximum Intensity Projection* visualization modes) and a three-dimensional, OpenGL-based isosurface viewer that can be zoomed and rotated, so that the radiologist can have a better understanding of the local morphological structure (see Fig. 3). The parameters of the selected region (label, size, location of the center) can then be corrected according to these more precise observations.

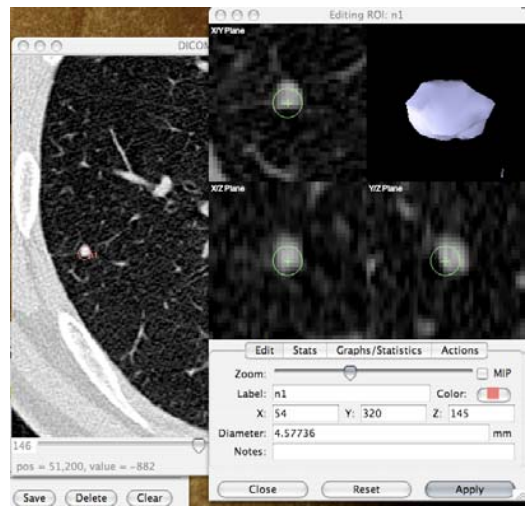

Fig. 3. The detailed ROI examination window showing orthogonal cross-section planes and a three-dimensional isosurface rendering.

The tool is free software released under the GNU Public License, it is written in C++ and it takes advantage of the open source QT GUI toolkit[4] for the graphical interface. It was written to work on the most common operating systems, and it has been successfully tested on Unix/Linux, Mac OS, and Windows (using the Cygwin POSIX-compatibility layer). The code structure is modular, in an effort to simplify extension and modification of the basic behaviors, as well as giving the possibility to add more complex schemes of interaction (like the contour identification that was previously mentioned), and, eventually, different data sources.

As 3D CT datasets tend to be very large, the tool performs heavy data caching, so that the memory footprint of the application is limited even when dealing with very large datasets: the full dynamic range images are always kept on disk, while the rendered (i.e. post-processed, 8-bit-mapped) images are cached in memory and generated in the background. This allows for fast slice navigation once the radiologist is satisfied with the display parameters.

This software has been in use for more than one year at the Department of Physics at the University of Pisa, where it has been used by radiologists to annotate more than 150 exams. Feedback from them has been very satisfactory.

### 2.3 The CAD strategy

Pulmonary nodules may be characterized by very low CT values and/or low contrast, may have CT values similar to those of blood vessels and airway walls or may be strongly connected to them (see Fig. 1).

The strategy we adopted focuses first on the detection of nodule candidates by means of a 3D enhancing filter emphasizing spherically-shaped objects. As a second step, the reduction of false-positive findings by means of a voxel-based neural approach is implemented. The two steps of the analysis are applied only once a purposely built segmentation algorithm has identified lung internal and sub-pleural regions, where the two types of nodules have to be searched. The two regions are defined according to extremely conservative criteria and they are partially overlapped (see Fig. 4).

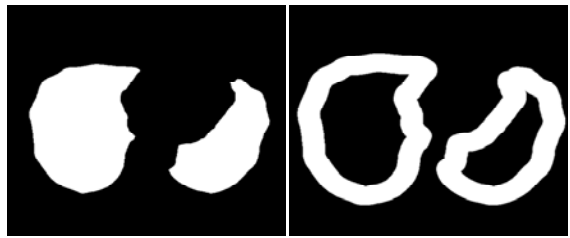

Fig. 4. Lung internal (left) and sub-pleural (right) regions identified by the segmentation algorithm.

The automated nodule candidate detection should be characterized by a sensitivity value close to 100%, in order to avoid setting an *a priori* upper bound to the CAD system performances. To this aim, lung nodules are modelled as spherical objects and a dot-enhancement filter is applied to the 3D matrix of voxel data. The filter attempts to determine the local geometrical characteristics of each voxel, by computing the eigenvalues of the Hessian matrix and evaluating a *magnitude* and a *likelihood* functions specifically configured to discriminate between the local morphology of linear, planar and spherical objects, the latest modelled as having 3D Gaussian sections[5,6]. To enhance the sensitivity of this filter to nodules of different sizes, a multi-scale approach was followed. According to the indications given in [5,7,8], a Gaussian smoothing at several scales was implemented. In particular, the Gaussian smoothing and the computation of the Hessian matrix were combined in a convolution between the original data and the second derivatives of a Gaussian smoothing function. This procedure is based on an *a priori* knowledge of the size range of the nodules to be enhanced. The range and the number of intermediate smoothing scales must be determined on the basis of the dataset of available CT scans. Once a set of N filtered images is computed, each voxel of the 3D space matrix is assigned the maximum *magnitude* x *likelihood* value obtained from the different scales, multiplied by the relative scale factor, according to [5,8]. A peak-detection algorithm is then applied to the filter output to detect the local maxima in the 3D space matrix. The final filter output is a list of nodule candidates sorted by the value the filter function assigned.

The dot-enhancement filter is run on the union of the lung internal and sub-pleural regions identified by the segmentation algorithm. Then, from the filter output list, two different lists of nodule candidates are created, depending on their location in one region or in the other one. Obviously the two lists can have some elements in common, due to the partial overlapping of the lung internal and sub-pleural regions. The lists in both internal and sub-pleural lung regions are then truncated so to keep the sensitivity close to 100%, therefore accepting a large number of false positives.

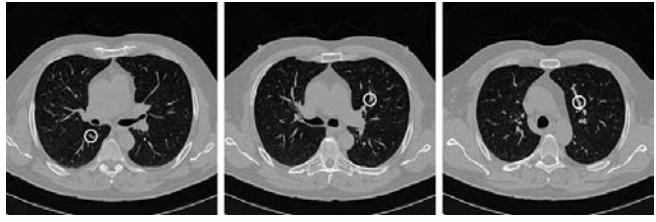

Fig. 5. Some examples of false positive findings generated by the dot-enhancement filter.

In Fig. 5 some examples of FP findings are shown. To reduce the amount of FP/scan, we developed a procedure we called voxel-based neural approach (VBNA). Each voxel of a region of interest (ROI) is characterized by a feature vector constituted by the grey level intensity values of its 3D neighbors and the eigenvalues of the *gradient matrix D* defined by

$$D_{ij} = \sum \partial_{x_i} \partial_{x_j} \quad i,j = 1,2,3,$$

where the sums are over the neighborhood area, and the eigenvalues of the Hessian matrix $H$

$$H_{ij} = \partial^2_{x_i x_j} \quad i,j = 1,2,3$$

(see Fig. 6)[9]. As we proved in [10], the implementation of these six additional features exploiting the morphological properties of the voxel neighborhood improves the neural system discrimination power. A feed-forward neural network is trained and tested at this stage assigning each voxel either to the nodule or normal tissue target class. A candidate nodule is then characterized as "CAD nodule" if the number of voxels within its ROI tagged as "nodule" by the neural classifier is above some relative threshold. A free response receiver operating characteristic (FROC) curve for our CAD system can therefore be evaluated at different threshold levels.

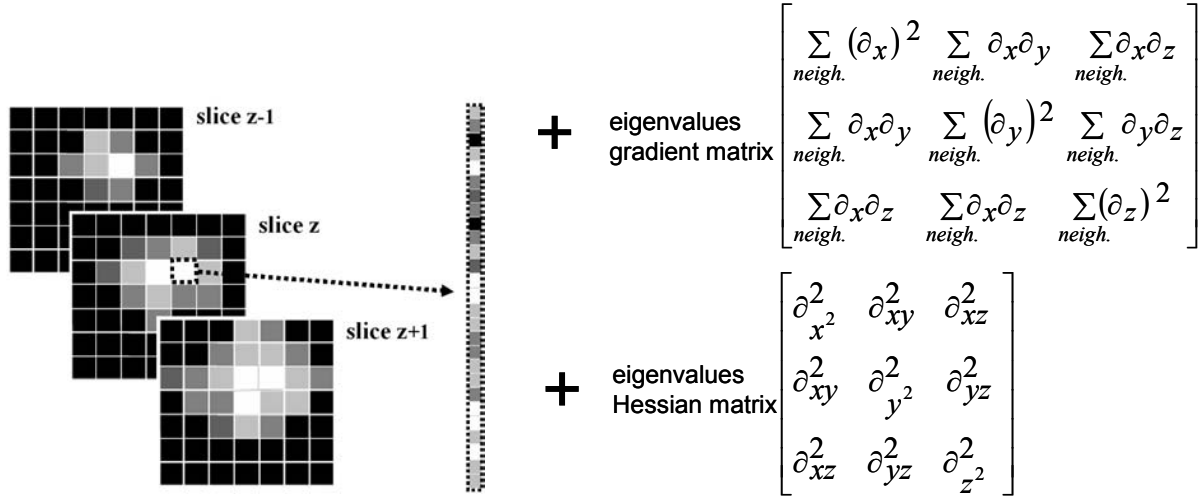

Fig. 6. Voxel-based neural approach to false-positive reduction: each voxel is characterized by a feature vector constituted by the grey level intensity values of its 3D neighbors and the eigenvalues of the gradient and the Hessian matrices.

# 3. DATA ANALYSIS AND RESULTS

The dataset used for this study consists of 39 CT scans, containing 75 internal nodules and 27 sub-pleural nodules with diameter greater than 5 mm; nodule sizes were estimated on the basis of radiologists' annotations. As shown in Fig. 7, the distribution of nodule diameters greater than 5 mm reaches the maximum value of about 12 mm, whereas most of nodules have diameter ranging from 5 to 8 mm.

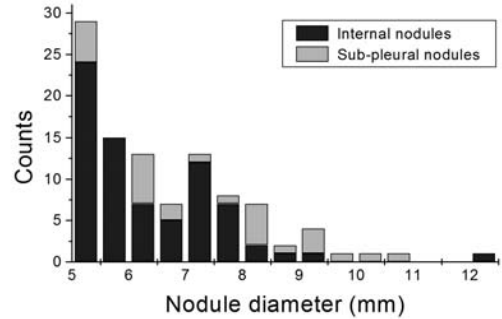

Fig. 7. Distribution of diameters of internal and sub-pleural nodules.

Out of the 39 available scans, only 34 contain internal nodules and only 20 contain sub-pleural nodules.

Once the lung internal and sub-pleural regions have been identified and the 3D dot-enhancement filter has provided the lists of nodule candidates in both the regions, the VBNA is applied to reduce the amount of FP findings.

## 3.1 Internal nodules

The available dataset of 34 scans containing 75 internal nodules was partitioned into a teaching set of 15 scans containing 30 nodules and a validation set of 19 scans containing 45 nodules; the partition was defined so that the teaching set is representative of all the nodule dimensions. Five three-layer feed-forward neural networks were trained on five different random partitions of the teaching set into train and test sets. The performances achieved in each trial for the correct classification of individual voxels are reported in Table 1, where the sensitivity and the specificity values obtained on the test sets and on the whole teaching set in the five trials are shown.

Table 1. VBNA internal nodules: performances achieved by five neural networks on five random partitions of teaching set into train and test sets

| Test set | | Teaching set | |
|---|---|---|---|
| sens % | spec % | sens % | spec % |
| 78.2 | 83.8 | 85.6 | 86.0 |
| 80.0 | 82.8 | 87.5 | 86.6 |
| 77.2 | 85.3 | 81.7 | 86.2 |
| 74.0 | 82.3 | 85.6 | 85.9 |
| 77.6 | 79.3 | 84.8 | 81.0 |

The first three networks in Table 1 provided the best performances on test sets. Among them, the second one achieved the best performance on the teaching set. So the VBNA approach has been applied to each ROI identified by the dot-enhancement filter by using this trained neural network. The FROC curve (see Fig. 8) was evaluated on the whole dataset of 34 scans containing internal nodules, constituted by the teaching set of 15 scans containing 30 nodules and the validation set of 19 scans containing 45 nodules. As shown in figure, a sensitivity of 85.3% at 6 FP/scan is measured. If the sensitivity value is decreased to 74.7% a rate of 3.8 FP/scan is obtained.

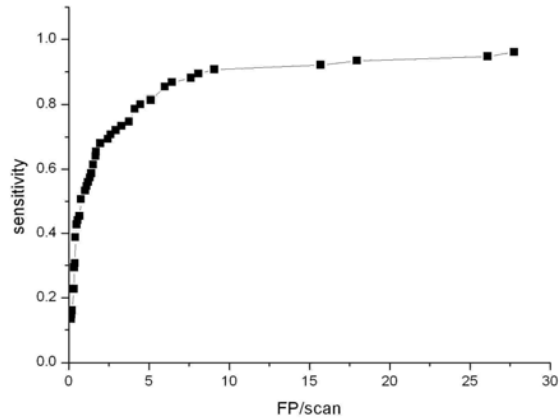

Fig. 8. FROC curve on the dataset of 34 scans containing 75 internal nodules.

**3.2 Sub-pleural nodules**

The same procedure of internal nodules was applied. In particular, the available dataset of 20 scans containing 27 sub-pleural nodules was partitioned into a teaching set of 17 scans containing 18 nodules and a validation set of 3 scans containing 9 nodules; the partition was defined so that the teaching set is representative of all the nodule dimensions. Five three-layer feed-forward neural networks were trained on five different random partitions of the teaching set into train and test sets. The performances achieved in each trial for the correct classification of individual voxels are reported in Table 2, where the sensitivity and the specificity values obtained on the test sets and on the whole teaching set in the five trials are shown.

Table 2. Sub-pleural nodules. VBNA: performances achieved by five neural networks on five random partitions of teaching set into train and test sets

| Test set | | Teaching set | |
|---|---|---|---|
| sens % | spec % | sens % | spec % |
| 73.0 | 91.1 | 83.7 | 92.6 |
| 75.9 | 89.3 | 82.5 | 91.3 |
| 61.0 | 90.8 | 76.9 | 91.6 |
| 72.6 | 85.7 | 80.0 | 84.6 |
| 63.5 | 88.3 | 78.8 | 90.0 |

The first two networks in Table 2 provided the best performances on test sets. Among them, the second one achieved the best performance on the teaching set. So the VBNA approach has been applied to each ROI identified by the dot-enhancement filter by using this trained neural network. The FROC curve (see Fig. 9) was evaluated on the whole dataset of 20 scans containing sub-pleural nodules, constituted by the teaching set of 17 scans containing 18 nodules and the validation set of 3 scans containing 9 nodules. As shown in figure, a sensitivity of 85.2% at 13.6 FP/scan is measured. If the sensitivity value is decreased to 74.1% a rate of 10.1 FP/scan is obtained.

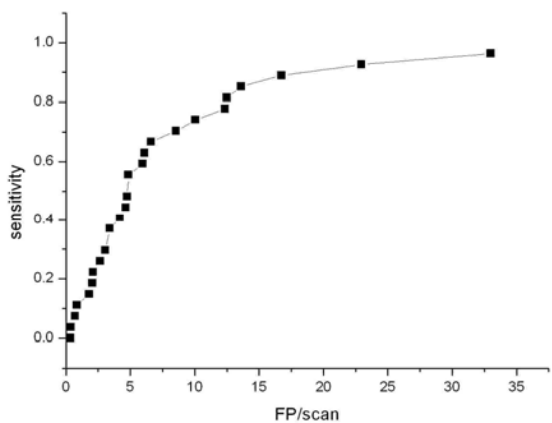

Fig. 9. FROC curve on the dataset of 20 scans containing 27 sub-pleural nodules.

It can be easily noticed that the CAD performance on the dataset of internal nodules (85.3% sensitivity at 6 FP/scan, 74.7% sensitivity at 3.8 FP/scan) is sensibly better than that achieved on sub-pleural nodules (85.2% sensitivity at 13.6 FP/scan, 74.1% sensitivity at 10.1 FP/scan). This behavior is clearly due to the sensibly different cardinality of the available datasets in the two cases.

## 4. CONCLUSIONS

The dot-enhancement pre-processing is a suitable tool for the identification of nodule candidates of diameter above 5 mm and the VBNA is an effective approach to the problem of false positives reduction. In particular the performance achieved by our CAD system in the detection of internal nodules is very satisfying: a very good sensitivity (85.3%) at a low level of false positives (6) is measured. The results obtained in the sub-pleural nodule detection are also promising (85.2% sensitivity at 13.6 FP/scan), but sensibly worse, due to the low number of sub-pleural nodules available for this study. A larger dataset of sub-pleural nodules is required to improve our CAD and make it an effective tool for both internal and sub-pleural nodule identification.

## ACKNOWLEDGEMENTS

This work was partially supported by Centro Studi e Ricerche Enrico Fermi and Bracco Imaging S.p.A.. We thank the researchers of the INFN- and MIUR-funded MAGIC-5 Collaboration for contributing to this work. We acknowledge Dr. L. Battolla, Dr. F. Falaschi and Dr. C. Spinelli of the U.O. Radiodiagnostica 2 dell'Azienda Ospedaliera Universitaria Pisana and Prof. D. Caramella and Dr. T. Tarantino of the Diagnostic and Interventional Radiology of the University of Pisa for providing the annotated database of CT scans. We are grateful to Dr M. Mattiuzzi from Bracco Imaging S.p.A. for useful discussions.

## REFERENCES

1.  S. Diederich, M.G. Lentschig, T.R. Overbeck, D. Wormanns, W. Heindel, "Detection of pulmonary nodules at spiral CT: comparison of maximum intensity projection sliding slabs and single-image reporting", *Eur. Radiol.* 11, 1345-1350 (2001).

2. www.cspo.it
3. http://root.cern.ch
4. http://www.trolltech.no/products/qt
5. Q. Li, S. Sone and K. Doi, "Selective enhancement filters for nodules, vessels, and airways walls in two- and three-dimensional CT scans", *Med. Phys*. 30(8), 2040-2051 (2003).
6. P. Delogu et al, "Preprocessing methods for nodule detection in lung *CT*", Computer Assisted Radiology and Surgery, Proceedings of the 19[th] International Congress and Exhibition (Berlin, Germany), *International Congress Series* 1281, 1099-1103 (2005).
7. J. Koenderink, "The structure of image", *Biol. Cybern*. 50, 363–370 (1984).
8. T. Lindeberg, "On scale detection for different operators", Proceedings of the Eighth Scandinavian Conference on Image Analysis, 857–866 (1993).
9. I. Gori, M. Mattiuzzi, "A method for coding pixels or voxels of a digital or digitalized image: Pixel Matrix Theory (PMT)", *European Patent Application* 05425316.6.
10. P. Delogu, M. E. Fantacci, I. Gori, A. Preite Martinez and A. Retico, "Computer-aided detection of pulmonary nodules in low-dose CT", Proceedings of the CompIMAGE Symposium, 20-21 Oct. 2006, Coimbra, Portugal.